\documentclass[runningheads]{llncs}
\usepackage[T1]{fontenc}
\usepackage{caption}
\usepackage{subfigure}
\usepackage{graphicx}
\usepackage{float} 
\usepackage{multirow} 
\usepackage{booktabs}
\usepackage{tabularx}
\usepackage[misc]{ifsym}
\usepackage{appendix}
\begin{document}
\title{Memory-efficient High-resolution OCT Volume Synthesis with Cascaded Amortized Latent Diffusion Models}
\titlerunning{Memory-efficient High-resolution OCT Volume Synthesis}
\author{Kun Huang\inst{1,4}, Xiao Ma\inst{1}, Yuhan Zhang\inst{2}, Na Su\inst{3}, Songtao Yuan\inst{3}, Yong Liu\inst{4}, Qiang Chen\inst{1}$^{(\textrm{\Letter})}$, and Huazhu Fu\inst{4}$^{(\textrm{\Letter})}$}
\institute{Nanjing University of Science and Technology \\ \email{chen2qiang@njust.edu.cn}
\and
The Chinese University of Hong Kong
\and
The First Affiliated Hospital with Nanjing Medical University
\and 
Institute of High Performance Computing, A*STAR
\\ \email{hzfu@ieee.org}
}
\maketitle              % typeset the header of the contribution
\begin{abstract}
Optical coherence tomography (OCT) image analysis plays an important role in the field of ophthalmology. Current successful analysis models rely on available large datasets, which can be challenging to be obtained for certain tasks. The use of deep generative models to create realistic data emerges as a promising approach. However, due to limitations in hardware resources, it is still difficulty to synthesize high-resolution OCT volumes. In this paper, we introduce a cascaded amortized latent diffusion model (CA-LDM) that can synthesis high-resolution OCT volumes in a memory-efficient way. First, we propose non-holistic autoencoders to efficiently build a bidirectional mapping between high-resolution volume space and low-resolution latent space. In tandem with autoencoders, we propose cascaded diffusion processes to synthesize high-resolution OCT volumes with a global-to-local refinement process, amortizing the memory and computational demands. Experiments on a public high-resolution OCT dataset show that our synthetic data have realistic high-resolution and global features, surpassing the capabilities of existing methods. Moreover, performance gains on two down-stream fine-grained segmentation tasks demonstrate the benefit of the proposed method in training deep learning models for medical imaging tasks. The code is public available at: \url{https://github.com/nicetomeetu21/CA-LDM}.

\keywords{Medical Image Synthesis \and Diffusion Probabilistic Model \and High-resolution Volumetric Images \and Memory-efficient Synthesis Framework.}
\end{abstract}
\section{Introduction}
Optical coherence tomography (OCT) has been widely used to visualize and exam the intricate retinal structure, with the advantages of non-invasive and high-resolution 3D volumetric imaging. By analyzing fine-grained pathological markers and anatomical changes within OCT volumes, numerous models has been developed to diagnose a variety of ocular and systemic diseases. 
For accurate and robust analysis, the development of these models requires the collection of sufficient OCT data under specific conditions. However, gathering such medical data frequently encounters many challenges, including the limitation of distribution of subjects, privacy concerns and reliance on expert knowledge.

Recent advancements in generative models, such as generative adversarial networks (GAN) \cite{creswell2018generative} and Diffusion models \cite{ho2020denoising}, show promising results in synthesizing high-quality images \cite{rombach2022high,saharia2022photorealistic} and videos \cite{blattmann2023align,chai2023stablevideo,esser2023structure}. These prompt researchers to explore the possibility of synthesizing medical images to address data-related challenges. However, prior works mainly focus on 2D images \cite{gao2023synthetic,zhou2020dr,10087263,jiang2020covid,ozbey2023unsupervised} or low-resolution 3D volumes. For instance, Khader et al. \cite{khader2023denoising} introduced a lightweight 3D latent diffusion model for the unsupervised synthesis of medical volumes, whereas Sun et al. \cite{sun2022hierarchical} developed a GAN-based hierarchical amortized framework. Deo et al. \cite{deo2023shape} employed conditional latent diffusion models with shape and anatomical guidance to synthesize 3D brain vasculatures. Hu et al. \cite{hu2022domain} explored transferring medical volumes to unseen domains by utlizing 2D variational autoencoders. Kim et al. \cite{kim2022diffusion} proposed a diffusion-based deformable model to estimate intermediate temporal volumes between source and target volumes.
Despite the high fidelity can be achieved, these methods are limited to synthesizing volumes of up to $256^3$ in size, due to the substantial memory requirements for the training and inference phases.
Since OCT volumes typically have higher resolutions, previous techniques could only produce down-sampled or cropped versions of OCT volumes, which lack critical fine-grained or long-term information. A recent study \cite{han2023medgen3d} proposes a sequence-based generation framework to generate high-resolution volumes slice-by-slice with available voxel-wise structural labels, which are expensive for OCT volumes. Therefore, there is a need for innovative methods to synthesize high-resolution OCT volumes, especially considering the limited memory capacity.

We introduce cascaded amortized latent diffusion models (CA-LDM) to synthesize high-resolution OCT volumes in a memory-efficient way. The core idea is to amortize the memory and computational demands of synthesizing whole OCT volumes to different diffusion processes in latent spaces. 
We first build a mapping between the low-resolution latent space and the high-resolution volume space with a group of autoencoder networks, so that the diffusion generation process can be taken in the low-resolution latent space. To address memory constraints, we introduce non-holistic autoencoders (NHAE). Unlike conventional autoencoders which encode and reconstruct the entire volume, NHAE performs volume super-resolution that involves encoding thumbnail volumes and then decoding them slice-by-slice into high-resolution volumes. In tandem with NHAE, cascaded diffusion processes are employed to synthesize high-quality OCT volumes. First, a 3D global diffusion process synthesizes 3D latent representations. Then, a 2D slice diffusion process refines the latent representations by injecting finer high-resolution details during the slice-wise sequential decoding. This design amortizes the requirements for depicting global and detailed features by different diffusion processes, thus the memory and computational demands are also amortized.
The proposed model could synthesize OCT volumes with the resolution of $512^3$, matching the inherent resolution of OCT volumes of many commercial devices.

The main contributions of our work are as follows: 1) Our proposed method is the first to synthesize 3D OCT volumetric images at a resolution of $512^3$, surpassing previously achieved resolutions. 2) We introduce non-holistic autoencoders for efficiently compressing and decompressing the high-resolution volumes into low-resolution representations. 3) We introduce cascaded diffusion processes for coarse-to-finely synthesizing high-quality volumetric images within the constraints of limited memory capacity. 4) Experimental results demonstrate the high quality of the generated volumetric images and the superiority compared existing methods. Besides, the results of two down-stream tasks highlight their potential benefits and applications.

\section{Methodology}
% \subsection{Preliminary}
% Denoising diffusion probabilistic model (DDPM) \cite{ho2020denoising} is a recent advanced image generation method. The forward process of DDPM transforms images x to standard Gaussian noises by $T$-steps noising. A U-shape denoising network is trained to estimate the added noise in the forward process and the diffusion process of DDPM generates images from a gaussian distribution by $T$-steps denoising. Based on DDPM, latent diffusion model (LDM) \cite{rombach2022high} is proposed to synthesize high-resolution images. LDM employs autoencoders to bidirectional map between image and latent space, and performs diffusion prediction in the latent space.

\subsection{Non-holistic Autoencoders}
\begin{figure}[t!]
\includegraphics[width=\textwidth]{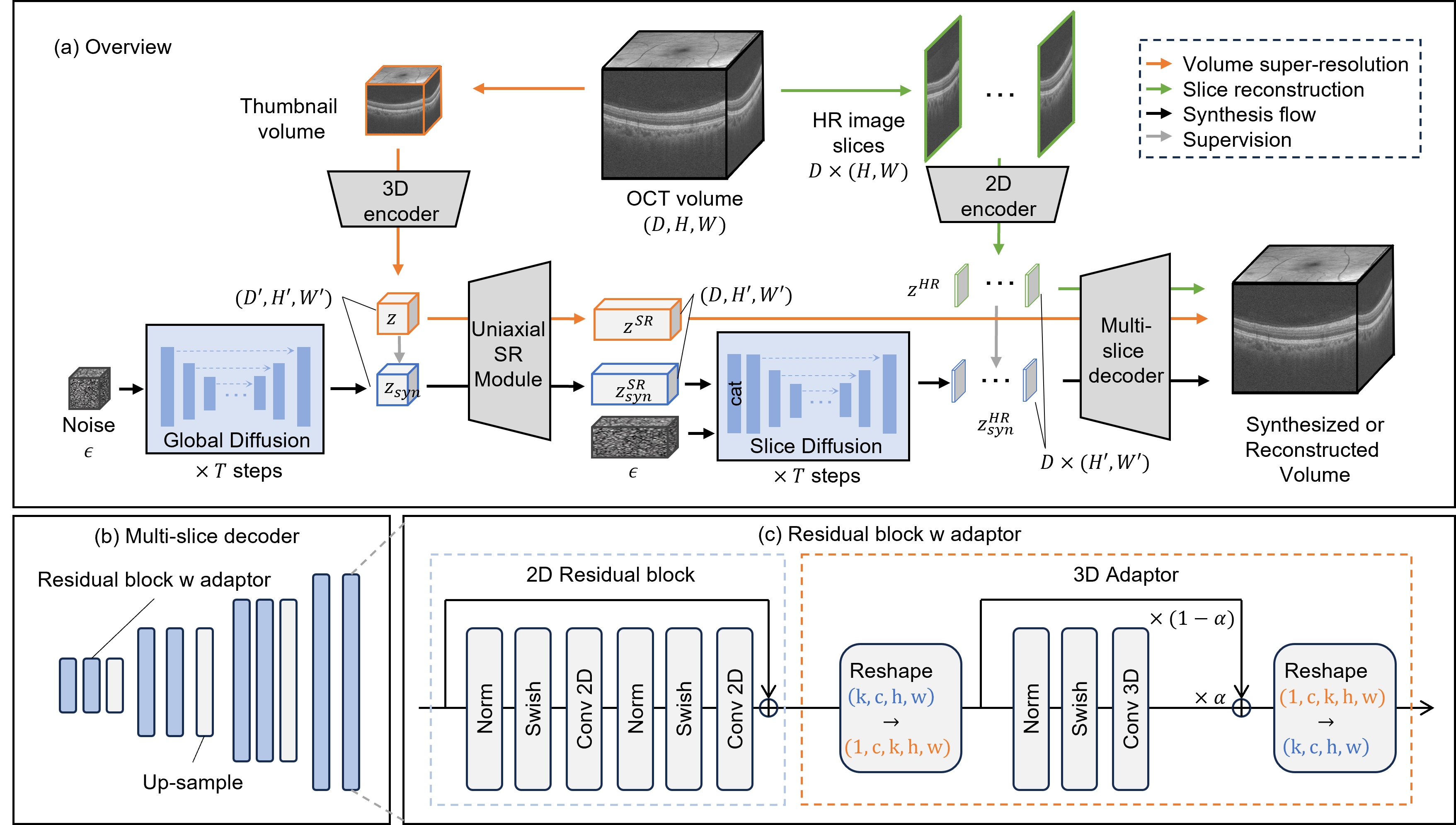}
\caption{(a) Overview of the proposed method. The size of images and latent representations are noted. (b) Architecture of the multi-slice decoder. It consists of 2D residual blocks with 3D adaptors of different scales. (c) Detailed architecture of the residual block and the 3D adaptor. $k,c,h,w$ represent the batch size, channels, height and width of a batch of 2D features. $\alpha$ is a learnable mixing factor.} \label{fig1}
\end{figure}

Fig.~\ref{fig1}(a) shows the overview of the proposed method. To synthesis 3D volumes, several previous works \cite{khader2023denoising,deo2023shape} migrate latent diffusion model (LDM) \cite{rombach2022high} to 3D versions. However, when the resolution of data becomes high, training the fully 3D autoencoders to encode and decode the entire volume is uneconomical. Instead, NHAE performs volume super-resolution with thumbnail image encoding, latent uniaxial super-resolution and a slice-wise sequential image decoding processes.

\subsubsection{Thumbnail image encoding:} 
We regard $x$ as the volumetric image of size $(D,H,W)$, where $D$ corresponds to the length of the inter-slice dimension and $H,W$ represent the height and width within the slices. In the encoding process, we first down-sample $x$ to a lower resolution thumbnail image. Then a fully 3D encoder takes the thumbnail image as input and outputs a latent representation $z$ of size $(c,D',H',W')$, where $c$ represent the number of channels.

\subsubsection{Latent uniaxial super-resolution:} We employ a uniaxial super-resolution module $f_{sr}$, consisting of multiple 3D residual and uniaxial upsample blocks, to up-sample the latent representation: $z^{sr}=f_{sr} (z)$. The up-sampled representation $z^{sr}$ has the size $(c,D,H',W')$ where the up-sampling is applied solely to the inter-slice dimension $D$.

\subsubsection{Slice-wise image decoding:}
We treat $z^{sr}$ as a sequence of the 2D latent slices of size $(c,H',W')$ and employ a multi-slice decoder that takes $k$ consecutive latent slices as input and outputs a center high-resolution image slice of size $(H,W)$. With a sequential decoding process, we can reconstruct the 3D volumetric image from $z^{sr}$ as follows:
\begin{equation}
    x^{rec}=[x_i]^D_{i=1},x_i=Dec_{M}([z^{sr}_j]^{i+\frac{k}{2}}_{j=i-\frac{k}{2}}) \label{eq1}
\end{equation}
where $x_i$ represents the $i$-th image slice of the reconstructed 3D volume $x^{rec}$, $z^{sr}_i$ represents the up-sampled 2D latent slice, and $Dec_M$ symbolizes the multi-slice decoder.

\subsubsection{Efficient multi-slice decoder:} Inspired by recent models [21,22] leveraging pretrained 2D networks to help model 3D data, we employ a 2D and 3D hybrid convolutional network as the multi-slice decoder to efficiently process the anisotropic 3D patches $(k\ll H',W')$. Fig.~\ref{fig1}(b)(c) illustrate the architecture of $Dec_{M}$. Specifically, we first train a 2D decoder that incorporates residual and up-sample blocks, to transform a single latent slice into its corresponding image slice. Then, we fix the trained parameters of the 2D networks and add learnable 3D convolutional adaptors after each residual block. When processing a batch of consecutive latent slices, the 2D layers operate on each slice independently, while the 3D adaptors reinterpret the batch dimension as a spatial dimension, executing 3D convolutions.

Both encoding and decoding of NHAE avoid direct process the entire high-resolution volume, reducing the memory requirement of the training and inference of networks with large parameters. For training, we randomly reconstruct one 2D slice image of each volumetric sample and calculate the same hybrid loss as LDM \cite{rombach2022high}.

\subsection{Cascaded Diffusion Processes}
Since NHAE bidirectional map between the volume space and the latent space, a global diffusion network $Diff_{3D}$ synthesizes 3D latent representations from  standard Gaussian noises $\epsilon$ by $T$-steps denoising \cite{ho2020denoising}, formulated as $z_{syn}=Diff_{3D}(\epsilon, T)$. Subsequently, we decode volumes from synthesized latent representations. Despite the high content consistency achieved by the synthetic volumes, some high-resolution details may be lost due to the high compression rate. To solve this problem, we introduce a slice diffusion refiner $Diff_{slice}$ to improve high-resolution details during the slice-wise decoding. Specifically, we further train an auxiliary 2D encoder alongside the fixed decoder. This encoder transforms high-resolution image slices $x_i$ into 2D latent slices $z^{hr}_{i}$ of size $(c,H',W')$. Compared to the 3D latent representation encoded from a down-sampled volume, $z^{hr}_i$ has more complete high-resolution information. The slice diffusion refiner learns to generate $z^{hr}_i$ guided by $z^{sr}_i$: $(z^{hr}_{syn})_i=Diff_{slice}(f_{sr}(z_{syn})_i,\epsilon, T)$. Then OCT volumes are decoded from the refined latent representations by Eq.~\ref{eq1}.

The global diffusion model is tasked with ensuring global consistency, whereas the slice diffusion refiner concentrates on enhancing high-resolution details, leveraging the accurate global context provided. This strategy not only optimizes resource use but also improves the quality of the synthesized high-resolution volumes.

% \begin{figure}[t!]
% \includegraphics[width=\textwidth]{fig2.png}
% \caption{(a) Architecture of the multi-slice decoder. It consists of 2D residual blocks with 3D adaptors of different scales. (b) Detailed architecture of the residual block and the 3D adaptor. $k,c,h,w$ represent the batch size, channels, height and width of a batch of 2D features. $\alpha$ is a learnable mixing factor.} \label{fig2}
% \end{figure}

\section{Experiments and Results}
\subsubsection{Dataset:} We conduct generation and downstream segmentation experiments on a public dataset OCTA-500 \cite{li2024octa}, which includes 300 high-resolution Optical Coherence Tomography (OCT) volumes of retinal macular area. As preprocess, we resize each OCT volumetric image from $400\times 640 \times 400$ to $512 \times 512 \times 512 (D \times H \times W)$. We randomly and category-individually split $2/3$ data (203 volumes) for training and $1/3$ data (97 volumes) for testing.

\subsubsection{Implementation details:} In our experiments, we set $T=1000$, the image size $D,H,W=512$, the latent size $D',H',W'=64$, $c=4$ and $k=5$. All experiments are implemented on 24G NVIDIA 4090 GPUs. To train CA-LDM, we first train NHAE, and then train two diffusion models. When inference by diffusion models, we employ the accelerate trick as the DDIM \cite{song2020denoising} that reduces the iterations to $T/5$. The synthesis speed is about 10 minutes per volumetric image. For label-guided synthesis of downstream segmentation tasks, we fine-tune diffusion models by adding condition encoders \cite{rombach2022high}. The encoded labels are concatenated with the noisy latent representations as the input of diffusion networks.

\begin{figure}[t!]
\includegraphics[width=\textwidth]{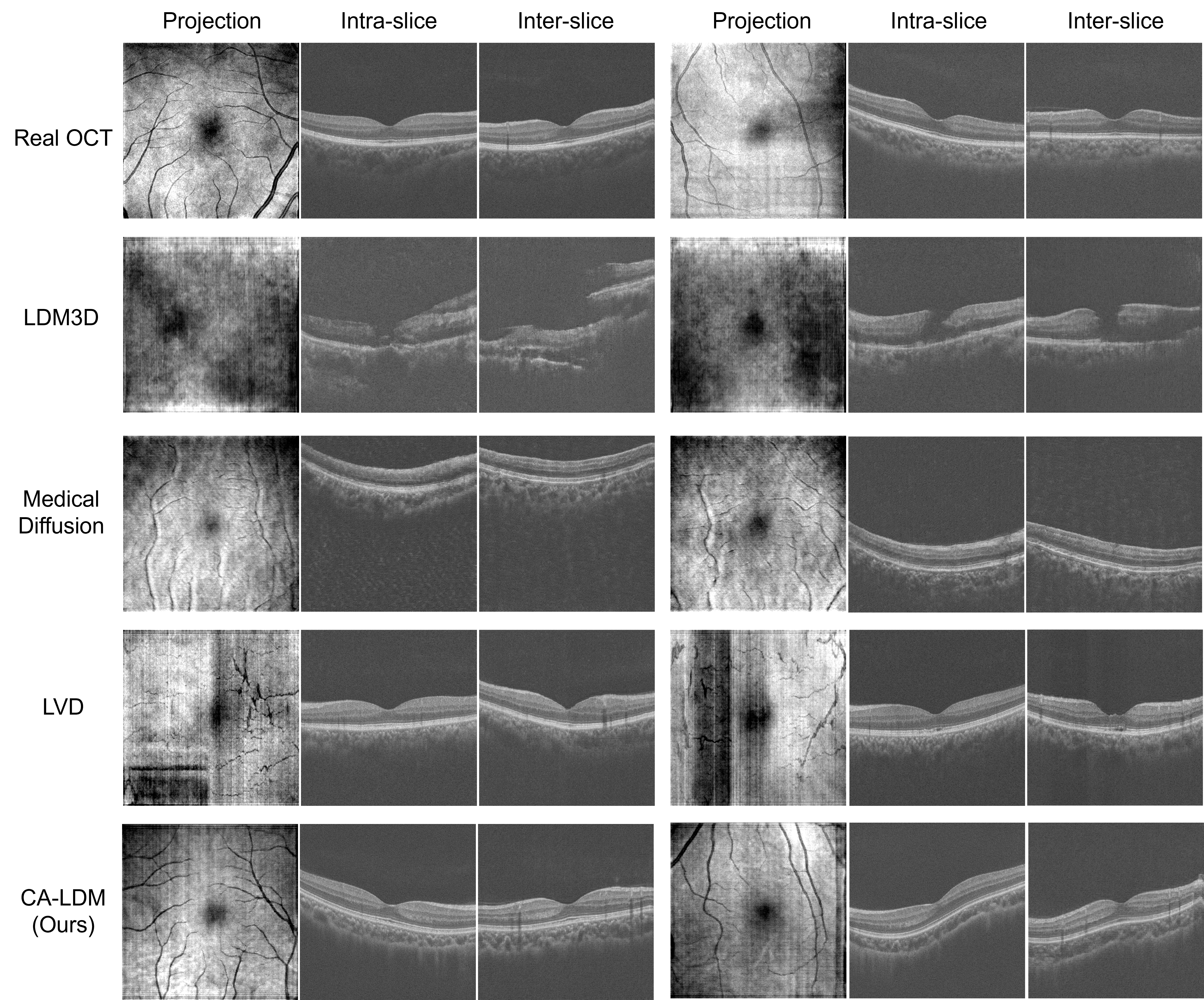}
\caption{Visual comparison of synthetic OCT. Each sample provides a mean projection of the whole volume and 2D images of intra-slice and inter-slice directions in the middle of the volume.} \label{fig3}
\end{figure}

\subsubsection{Metrics:} Fréchet inception distance (FID) \cite{heusel2017gans} and total variation (TV) \cite{rudin1992nonlinear} metrics are widely used to assess 2D image quality. A lower FID score indicates higher perceptual image quality. A lower TV score indicates cleaner and better visual quality. To evaluate the quality of the synthesized 3D volumes, we calculate TV of whole volumetric images, FID of all 2D images in intra-slice directions (Intra-FID), and FID of all 2D images in inter-slice directions (Inter-FID). For the downstream layer segmentation task, we calculate Dice and mean absolute distance (MAD) metrics \cite{zhang2021robust}. For the downstream artery–vein segmentation task, we calculate Intersection over Union metrics of arteries (IoU-A) and veins (IoU-V) \cite{li2024octa}.

\subsubsection{Comparison experiments:} 
\begin{figure}[t!]
    \begin{minipage}[t]{0.485\linewidth}
        \centering
        \includegraphics[width=\textwidth]{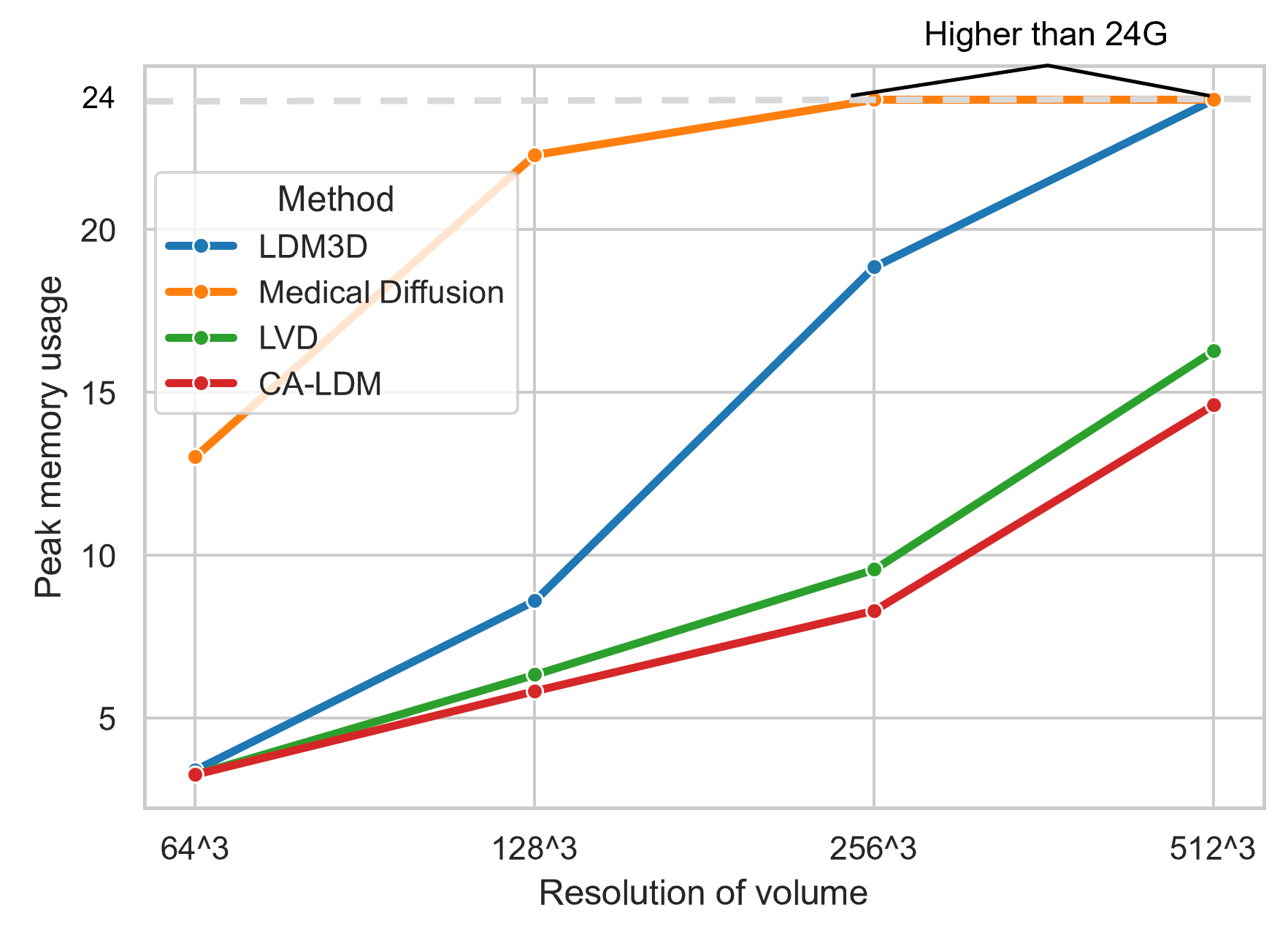}
        \caption{Peak memory usage during inference time with respect to the resolution of the synthesized volume for each model. These models are standard versions without parameter adjustments.} \label{fig5}
    \end{minipage}%
    \hspace{0.1in}
    \begin{minipage}[t]{0.485\linewidth}
        \centering
        \includegraphics[width=\textwidth]{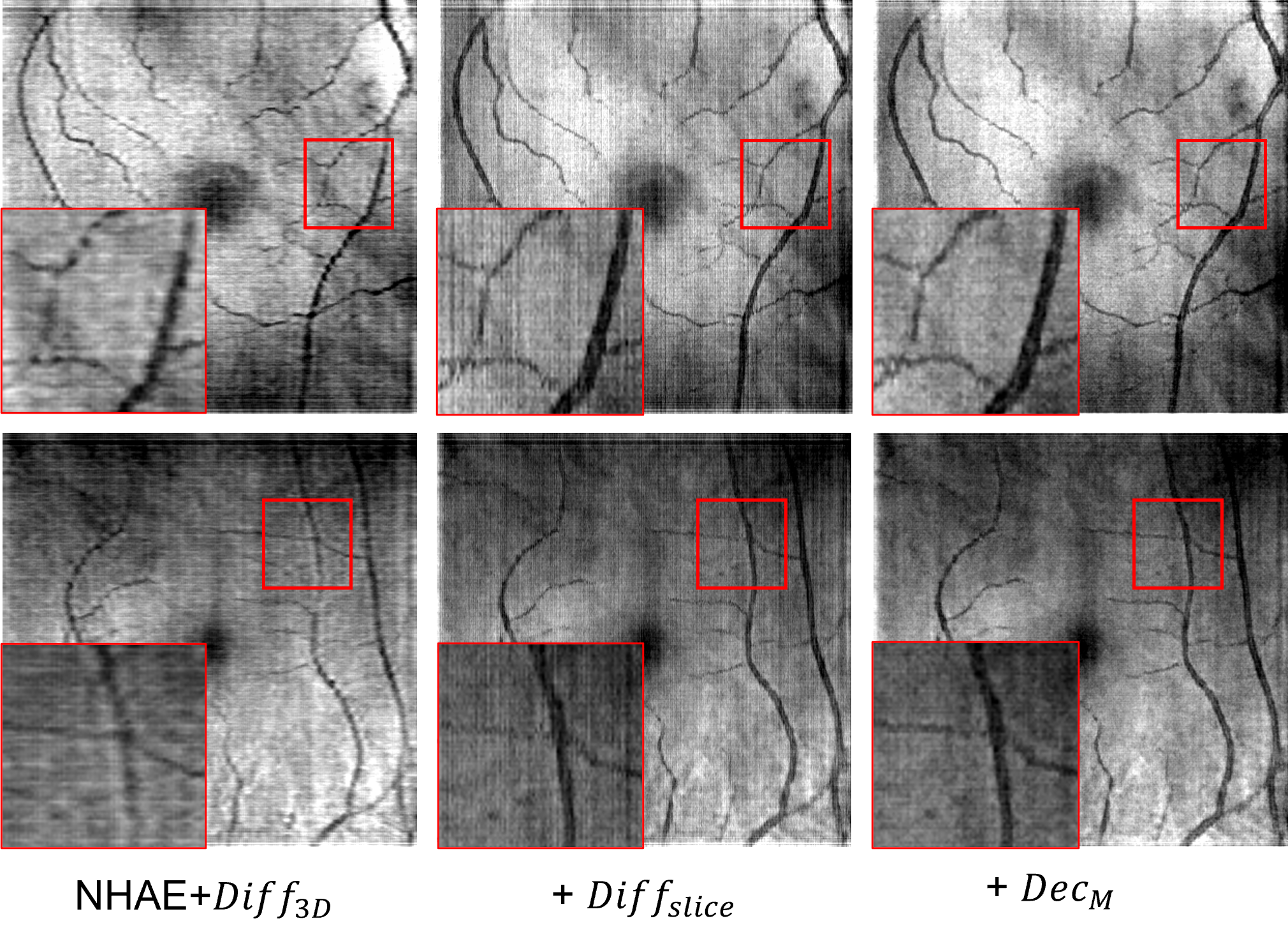}
        \caption{Projections of the ablation methods. Each group of samples is corresponding the same latent representations synthesized by $Diff_{3D}$.}  \label{fig4}
    \end{minipage}
\end{figure}

We compared the proposed CA-LDM with three 3D data synthesis models. These include a 3D version of LDM \cite{rombach2022high} (LDM3D) which employs 2D autoencoders coupled with a 3D diffusion model, Medical Diffusion \cite{khader2023denoising}, latent video diffusion (LVD) \cite{blattmann2023align}. The parameters of these models are adjusted to enable successful training and inference on 24G GPUs. Fig.~\ref{fig3} shows synthetic samples of different methods. The results of LDM3D exhibit incorrect structures and a notable lack of detail. Meanwhile, the results of Medical Diffusion \cite{khader2023denoising} lack high-resolution details and present noisy backgrounds. The results of LVD \cite{blattmann2023align} suffer from inconsistencies in global content, as evidenced by unauthentic vascular structures and brightness variations. In contrast, the results of CA-LDM have better high-resolution details and more consistent global content, making them the most similar to real OCT images. Table~\ref{tab1} shows the quantitative performance of each model. Our model demonstrates better Intra-FID, Inter-FID and TV scores. Both qualitative and quantitative assessments demonstrate that our proposed model delivers the highest quality of synthetic imagery, capable of producing authentic high-resolution volumetric images.
To demonstrate the efficient of CA-LDM in terms of memory usage, we monitor the peak memory usage of each model when synthesizing different resolution's volumes. As shown in Fig.~\ref{fig5}, CA-LDM has the lowest peak memory usage for each resolution's synthesis and the slowest growth rate when the resolution increasing.

\begin{table}[t!]
\centering
\caption{Quantitative evaluation of compared and ablation methods}\label{tab1}
\begin{tabular*}{\textwidth}{@{}@{\extracolsep{\fill}}l|c|c|c}
\toprule
Model &  Intra-FID $\downarrow$ & Inter-FID $\downarrow$ & TV $\downarrow$\\
\midrule
LDM3D \cite{rombach2022high} &  97.62 & 127.51 & 712.2 $\pm$ 14.0\\
Medical Diffusion \cite{khader2023denoising} & 48.90 & 54.60 & 639.9 $\pm$ 4.2\\
LVD \cite{blattmann2023align} & 21.78 & 54.57 & 654.9 $\pm$ 31.2\\
\midrule
NHAE+$Diff_{3D}$ & 39.06 & 52.09 & 609.4 $\pm$ 31.2 \\
NHAE+$Diff_{3D}$+$Diff_{slice}$ & 17.69 & 30.62 & 603.7 $\pm$ 13.2\\
NHAE+$Diff_{3D}$+$Diff_{slice}$+$Dec_{M}$ (CA-LDM) & \bf{17.10} & \bf{28.31} & \bf{575.3 $\pm$ 7.5}\\

\bottomrule
\end{tabular*}
\end{table}

% \begin{figure}[t!]
% \includegraphics[width=\textwidth]{fig4.png}
% \caption{Projections of the ablation methods. Each group of samples is corresponding the same latent representations synthesized by $Diff_{3D}$.} \label{fig4}
% \end{figure}

% \begin{figure}[t!]
% \includegraphics[width=\textwidth]{fig4.png}
% \caption{Projections of the ablation methods. Each group of samples is corresponding the same latent representations synthesized by $Diff_{3D}$.} \label{fig5}
% \end{figure}
% \begin{figure}
%  % \vspace{-0.8em}
%   \centering
% \subfigure[Projections of the ablation methods. Each group of samples is corresponding the same latent representations synthesized by $Diff_{3D}$.]{
%    \label{fig4}
%   \includegraphics[scale = 0.24]{fig4.png}
%   }
%    % \vspace{-0.6em}
% \subfigure[Projections of the ablation methods. Each group of samples is corresponding the same latent representations synthesized by $Diff_{3D}$.]{
%   \label{fig5}
%   \includegraphics[scale = 0.24]{fig4.png}
%   }
%   % \vspace{-0.6em}
% \caption{Experiment results for different unseen relation types on Re-DocRED and DocRED datasets.} 
%   % \label{fig:7} 
%    % \vspace{-1 em}
% \end{figure}

\subsubsection{Ablation study:} We perform ablation studies to evaluate the effectiveness of each proposed component, as shown in Table 1. Compared to the models with 2D autoencoders (LDM3D) and 3D autoencoders (Medical Diffusion), the model employing NHAE (NHAE+$Diff_{3D}$) has the lower Intra-FID, Inter-FID and TV scores. Each addition of $Diff_{slice}$ and $Dec_{M}$  leads to decreasing of all three scores. Besides, we show the synthetic volumes of each ablation model in Fig.~\ref{fig4}. It can be seen that incorporation of $Diff_{slice}$ improves the high-resolution details, but results still show inconsistency between slices, manifesting as the bright and dark strip patterns. Incorporation of $Dec_{M}$ mitigated the irregular patterns, resulting in a smoother synthetic volume.

\subsubsection{Benefits for downstream tasks:} We utilize CA-LDM to synthesize images guided by real labels for two 3D fine-grained segmentation tasks. One task is the artery–vein segmentation. We train the IPNv2 \cite{li2024octa} model to segment the retinal artery and vein vessels. Another is the layer segmentation. We train a modified 3D U-Net model \cite{zhang2021robust} to segment six retinal layer surfaces. The results are shown in Table 2. The first three rows in Table 2 shows that as the resolution decreases, the performance of the models also decrease, which demonstrates the importance of high-resolution data for OCT image analysis. Compared to the models of the "Real" group, models of the "200\% Syn" group have a higher IoU-V and a lower MAD scores, which demonstrates the potential of synthetic data to replace the real data. Besides, models of the "Real+100\% Syn" group have better metrics of both tasks than model of "Real" group. And the inclusion of an additional 100\% synthetic data (Real+200\% Syn) leads to further improvements in performance for both tasks. These demonstrate the synthetic data can supplement real data to benefit downstream tasks. The qualitative results and synthetic data are shown in Appendix.

\begin{table}[t!]
    % \scriptsize
    \centering
    \caption{Segmentation performance of two tasks using real data (Real), real data with a downsampled resolutions of $128^3$ and $256^3$ (Real-128,Real-256), twice the amount of synthetic data (200\% Syn), real and an equivalent amount of synthetic data (Real+100\% Syn), real and twice the amount of synthetic data (Real+200\% Syn).}
    
    \label{tab:2}
    \begin{tabular*}{\textwidth}{@{}@{\extracolsep{\fill}}l|c|c|c|c@{}}
        \toprule
        \multirow{2}{*}[-0.5ex]{Training data} & \multicolumn{2}{c|}{Artery–vein Seg} & \multicolumn{2}{c}{Layer Seg}  \\
        \cmidrule(r){2-5}  
        &	IoU-A $\uparrow$ & IoU-V $\uparrow$  & Dice $\uparrow$ &	MAD $\downarrow$\\
        \midrule
        Real-128 & 0.427 $\pm$ 0.048 & 0.425 $\pm$ 0.064 & 0.884 $\pm$ 0.057 & 1.647 $\pm$ 1.492 \\
        Real-256 & 0.589 $\pm$ 0.054 & 0.585 $\pm$ 0.585 & 0.918 $\pm$ 0.071 & 1.216 $\pm$ 2.167 \\
        Real & 0.681 $\pm$ 0.058 & 0.677 $\pm$ 0.091 & 0.941 $\pm$ 0.075 & 1.030 $\pm$ 2.032\\
        200\% Syn  & 0.679 $\pm$ 0.059 & 0.679 $\pm$ 0.091 & 0.941 $\pm$ 0.075 & 0.976 $\pm$ 1.723\\
        Real+100\% Syn & 0.681 $\pm$ 0.059 & 0.679 $\pm$ 0.090 & 0.943 $\pm$ 0.075 & 0.931 $\pm$ 1.882\\
        Real+200\% Syn & \bf{0.683 $\pm$ 0.063} & \bf{0.680 $\pm$ 0.099} & \bf{0.944 $\pm$ 0.071} & \bf{0.930 $\pm$ 1.894}\\
        \bottomrule
    \end{tabular*}
\end{table}

\section{Conclusion}
This paper presents a cascaded amortized latent diffusion model (CA-LDM) to synthesize high-resolution medical volumetric images. Addressing the challenges posed by restricted memory resources, we propose non-holistic autoencoders and cascaded diffusion processes. The experiments demonstrate that our proposed method outperform state-of-the-art methods and is helpful for downstream segmentation tasks. Future efforts will focus on minimizing the time consumption of the model and achieving more flexible control over the synthesis process.
\begin{credits}
\subsubsection{\ackname} 
This work was supported in part by  the Chinese Scholarship Council; the Major Research Plan of the National Natural Science Foundation of China under Grant (92370109,
62202408); National Natural Science Foundation of China (62172223, 61671242); Fundamental Research Funds for the Central Universities
(30921013105); the Huazhu Fu’s Agency for
Science, Technology and Research (A*STAR) Career Development Fund (C222812010) and Central Research Fund (CRF).
\end{credits}
\bibliographystyle{splncs04}
\bibliography{mybibliography}
%





% LNCS Homepage, \url{http://www.springer.com/lncs}, last accessed 2023/10/25
% \end{thebibliography}
\appendix
\setcounter{table}{0}   %从0开始编号，显示出来表会A1开始编号
\setcounter{figure}{0}
%定义编号格式，在数字序号前加字符“A"
\renewcommand{\thetable}{A\arabic{table}}
\renewcommand{\thefigure}{A\arabic{figure}}
\begin{figure}
\includegraphics[width=\textwidth]{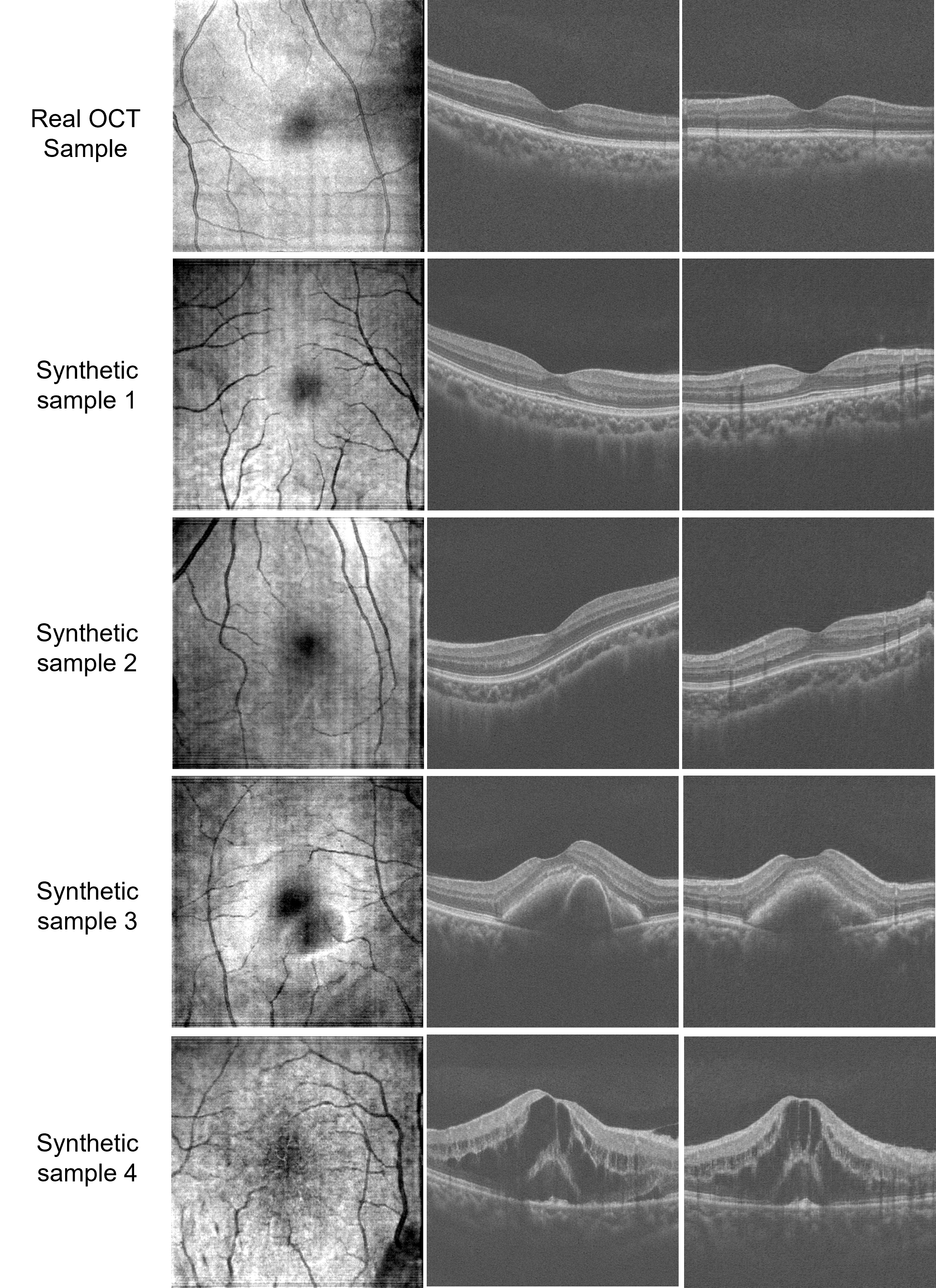}
\caption{High-resolution synthetic samples of CA-LDM. The first three rows are corresponding to samples in Fig. 2. The last two rows are samples with obvious pathological features.} \label{figS1}
\end{figure}

\begin{figure}
\includegraphics[width=\textwidth]{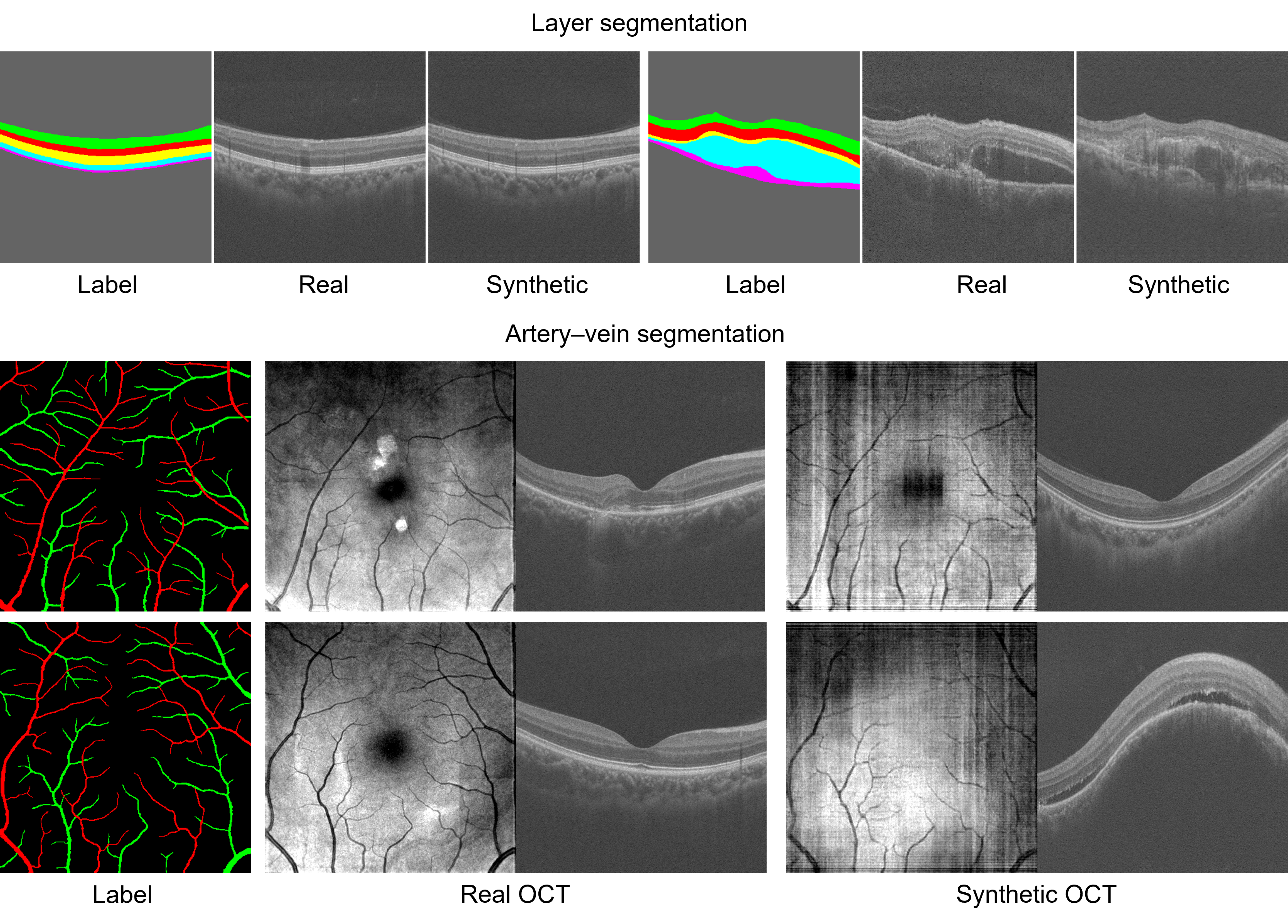}
\caption{Samples of label guided synthesis of CA-LDM for two tasks, given as projection and slice views. It can be found that CA-LDM has the capability to achieve feature disentanglement, enabling the generation of more varied image-label pairs.} \label{figS2}
\end{figure}

\begin{figure}
\includegraphics[width=\textwidth]{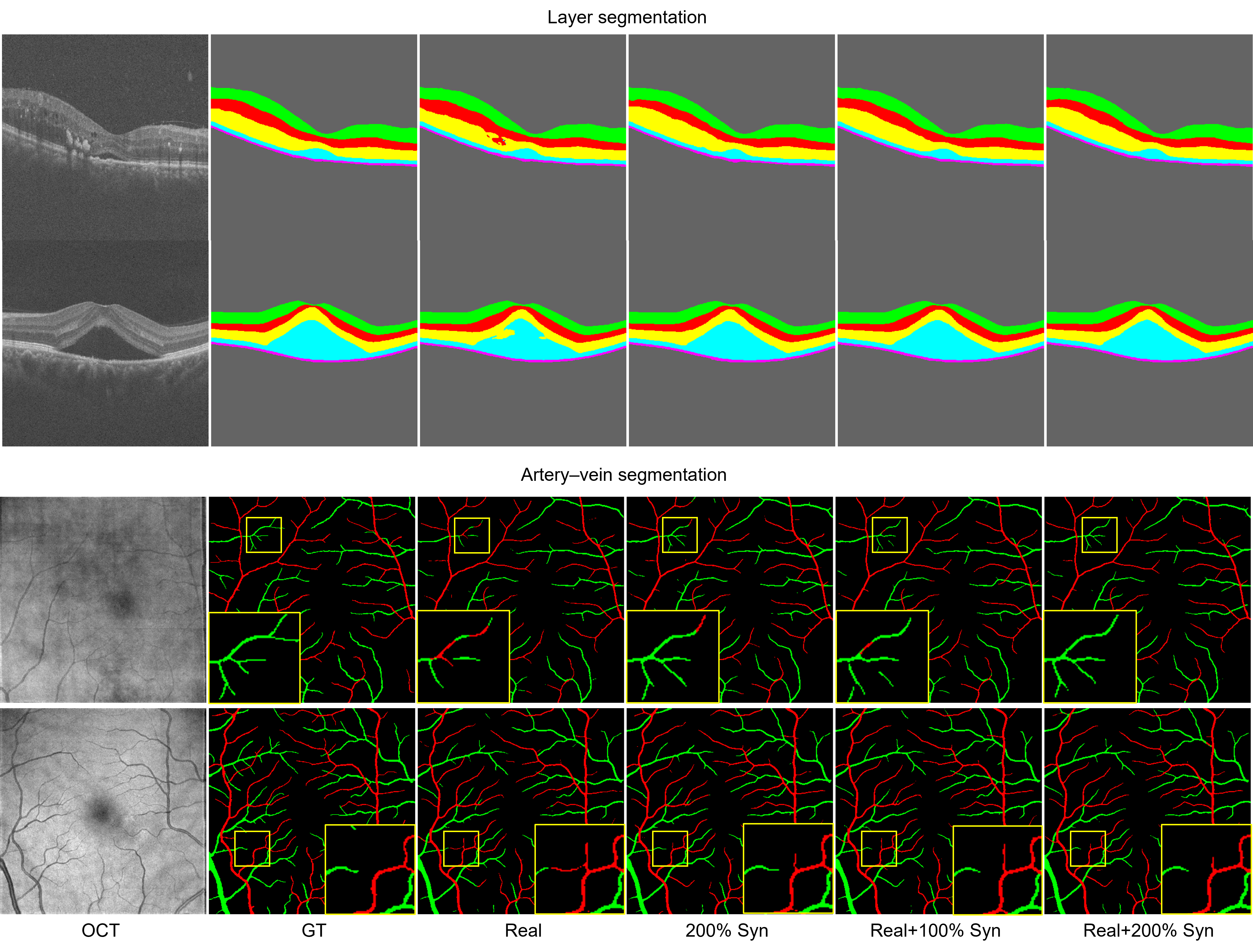}
\caption{Qualitative results of models trained by different data corresponding to Table. 2.} \label{figS3}
\end{figure}

\end{document}